\documentclass[journal,twoside,twocolumn,a4paper]{IEEEtran}

\usepackage{amsmath,amssymb}
\usepackage[dvips,colorlinks=true,bookmarks=false,citecolor=blue,urlcolor=blue]{hyperref} 

\usepackage{graphicx,wrapfig,balance}
\usepackage{psfrag}

\newcommand{\Cfs}{\mathcal{C}_{4,16}}
\newcommand{\Cft}{\mathcal{C}_{4,256}}

\newcommand{\bU}{\textit{\textbf{U}}}
\newcommand{\bX}{\textit{\textbf{X}}}\newcommand{\bx}{\textit{\textbf{x}}}
\newcommand{\bY}{\textit{\textbf{Y}}}
\newcommand{\bZ}{\textit{\textbf{Z}}}
\newcommand{\mcX}{\mathcal{X}}
\newcommand{\set}[1]{\{#1\}}
\newcommand{\ie}{i.e.,~}

\newcommand{\ld}{\ldots}

\newcommand{\Eb}{E_\tr{b}}
\newcommand{\No}{N_{0}}
\newcommand{\Es}{E_\tr{s}}
\newcommand{\tr}[1]{\mathrm{#1}}

\newcommand{\un}[1]{\underline{#1}}
\newcommand{\figref}[1]{Fig.~\ref{#1}}

\newcommand{\Rc}{R_\tr{c}}
\newcommand{\R}{R}
\newcommand{\MODwoa}{\Phi}      		

\usepackage{multicol}

\title{Achievable Rates for Four-Dimensional Coded Modulation with a Bit-Wise Receiver}
\author{Alex Alvarado and Erik Agrell
\thanks{A.~Alvarado is with the Dept.~of Engineering, University of Cambridge, Cambridge CB2 1PZ, United Kingdom (email: alex.alvarado@ieee.org).}
\thanks{E.~Agrell is with the Dept.~of Signals and Systems, Chalmers Univ.~of Technology, SE-41296 G\"oteborg, Sweden (email: agrell@chalmers.se).}
}

\begin{document}
\maketitle

\begin{abstract}
We study achievable rates for four-dimensional (4D) constellations for spectrally efficient optical systems based on a (suboptimal) bit-wise receiver. We show that PM-QPSK outperforms the best 4D constellation designed for uncoded transmission by approximately 1 dB. Numerical results using LDPC codes validate the analysis.
\end{abstract}

\section{Introduction}\label{Sec:Introduction}

Long-haul fiber-optic communication systems utilize coherent transmission, where both quadratures and both polarizations of the electromagnetic field are used, resulting in a four-dimensional (4D) signal space. To meet the demands for spectral efficiency, multiple bits are encapsulated in each constellation symbol. Combined with the use of forward error correction (FEC), this leads to the challenging problem of designing coded modulation (CM) schemes for optical communications \cite{Bulow2011}. 

One promising candidate for future optical CM systems is the (noniterative) bit-interleaved coded modulation (BICM) paradigm \cite{Zehavi92,Caire98}. The key feature of BICM is a suboptimal but flexible \emph{bit-wise receiver}. In BICM, the detection process is decoupled: soft information on the bits is calculated and then a soft FEC decoder is used. BICM systems are used in most wireless standards and they have also been considered for optical communications in, \ie \cite{Bulow2011,Djordjevic2007_JLT,Batshon2009_JLT}.

From an information-theoretic point of view, the suboptimality of a bit-wise receiver is reflected in terms of achievable rates, i.e., in the number of bits per symbol that can be reliably transmitted through the channel. The mutual information (MI) is the largest achievable rate for any kind of communication scheme. For a bit-wise decoder, this quantity is replaced by the so-called generalized mutual information (GMI). Although the MI and the GMI coincide when the signal-to-noise ratio (SNR) tends to infinity, the MI is strictly larger than the GMI for any finite SNR. This penalty depends on the constellation and its binary labeling and can be very large \cite[Fig.~4]{Caire98}, \cite{Agrell2011}.

In this paper, we study achievable rates for 4D constellations with a bit-wise receiver. The results show that constellations that are good for uncoded systems are also good in terms of MI. However, these constellation are not necessarily the best choice for coded systems based on (suboptimal) a bit-wise receiver. Numerical results based on low-density parity check (LDPC) codes confirm the theoretical analysis.

\section{Model and Achievable Rates}

We consider the vectorial additive white Gaussian noise (AWGN) channel $\bY=\bX+\bZ$, where $\bX, \bY, \bZ$ are four-dimensional real vectors. The transmitted vector $\bX$ is selected with equal probability from a constellation $\mcX\triangleq\set{\bx_1,\bx_2,\ld,\bx_{M}}$, where $M=2^m$. The components of the noise vector $\bZ$ are independent, zero-mean, Gaussian random variables with variance $\No/2$ in each dimension. The average symbol energy is $\Es\triangleq (1/M)\sum_{i=1}^{M}\|\bx_{i}\|^2$. 

\begin{figure}[ht]
\begin{center}
\newcommand{\scale}{0.86}
\psfrag{bi}[br][Br][\scale]{$\un{\bU}$}
\psfrag{CHE}[cc][cc][\scale]{ENC}
\psfrag{bc1}[cc][cc][\scale]{$\un{B}_{1}$}
\psfrag{bcm}[cc][cc][\scale]{$\un{B}_{m}$}
\psfrag{MOD}[cc][cc][\scale]{$\MODwoa$}
\psfrag{ENCODER}[Bc][Bc][\scale]{CM transmitter}
\psfrag{x}[bc][Bc][\scale]{$\un{\bX}$}
\psfrag{ddd}[cc][cc][\scale]{$\vdots$}
\psfrag{y}[bc][Bc][\scale]{$\un{\bY}$}
\psfrag{ML}[Bc][Bc][\scale]{Optimum receiver}
\psfrag{DECML}[cc][cc][\scale]{ML receiver}
\psfrag{BI}[Bc][Bc][\scale]{Bit-wise receiver}
\psfrag{MOD2}[cc][cc][\scale]{$\MODwoa^{-1}$}
\psfrag{lc1}[cc][cc][\scale]{$\un{L}_{1}$}
\psfrag{lcm}[cc][cc][\scale]{$\un{L}_{m}$}
\psfrag{DEC}[cc][cc][\scale]{DEC}
\psfrag{bih}[bl][Bl][\scale]{$\un{\hat{\bU}}$}
\psfrag{(a)}[Bc][Bc][\scale]{(a)}\psfrag{(b)}[Bc][Bc][\scale]{(b)}\psfrag{(c)}[Bc][Bc][\scale]{(c)}
\includegraphics[width=0.65\columnwidth]{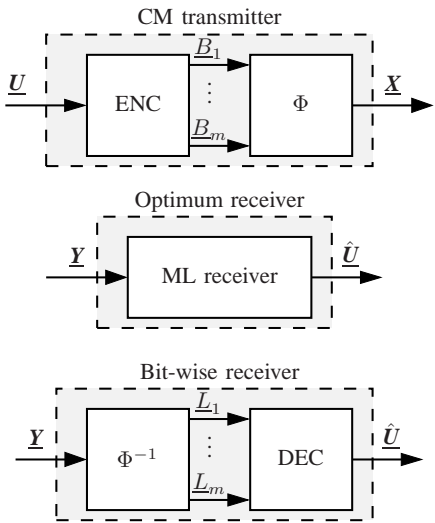}
\end{center}
\caption{Transmitter and receiver structures.}
\label{CM_Model}
\end{figure}

\figref{CM_Model} shows the considered transmitter and receiver structures. The CM transmitter consists of a FEC encoder (ENC), which encodes a binary input sequence $\un{\bU}$ into $m$ binary sequences $\un{B}_1, \ldots, \un{B}_m$, and a memoryless mapper $\MODwoa$, which maps $\un{B}_1, \ldots, \un{B}_m$ into a sequence of symbols $\un{\bX}$, one symbol at a time. For a rate $\Rc$ FEC encoder, the transmission rate in bit/symbol is $\R=\Rc m$. The average bit energy is $\Eb=\Es/\R$.  At the receiver side, an optimal receiver based on the maximum-likelihood (ML) rule can be implemented. An alternative to the ML receiver is a (suboptimal) bit-wise receiver. In this case, soft information on the coded bits $\un{B}_1, \ldots, \un{B}_m$ is calculated by $\MODwoa^{-1}$, typically in the form of logarithmic likelihood ratios (LLRs) $\un{L}_1, \ldots, \un{L}_m$. These LLRs are then passed to an off-the-shelf soft FEC decoder (DEC).\footnote{Alternatively, a hard-decision demapper can be combined with a binary-input decoder.} The bit-wise receiver in \figref{CM_Model} is usually known as a BICM receiver, owing its name to the original works \cite{Zehavi92,Caire98}, where a bit-level interleaver was included between the encoder and mapper. We refrain from using such a name because the interleaver might or might not be included, and if included, we assume it to be part of the FEC encoder.\footnote{Note that when an interleaver is included, ML decoding becomes impractical, and thus, the bit-wise receiver is the preferred alternative.}

For the optimum ML receiver and a given constellation $\mcX$, the largest achievable rate is the MI between $\bX$ and $\bY$, denoted by $I(\bX;\bY)$. Thus, decoding with arbitrarily low probability of error is possible if $\R\leq I(\bX;\bY)$. On the other hand, an achievable rate for the bit-wise receiver in \figref{CM_Model} is the GMI, given by $\sum_{k=1}^{m}I(B_{k};\bY)$, where $B_{k}$ is the $k$th bit at the mapper's input (see \figref{CM_Model}). It follows from the chain rule of MI that $\sum_{k=1}^{m}I(B_{k};\bY) \leq I(\bX;\bY)$, which can be understood as the loss in terms of achievable rates caused by the use of a bit-wise receiver. Furthermore, the GMI (unlike the MI) is highly dependent on the binary labeling (\ie $\Phi$ in \figref{CM_Model}). Gray codes are known to be good for high SNR \cite[Fig.~4]{Caire98}, \cite{Agrell2011}, \cite[Sec.~IV]{Alvarado11b}, but for many constellations, they do not exist.

\section{Numerical Results}\label{Sec:NumResults}

We consider three 4D constellations with $M=16$ ($m=4$): polarization-multiplexing quadrature phase-shift keying (PM-QPSK) \cite{Sun2008_OE}, the constellation $\Cfs$ introduced in \cite{Karlsson2010}, which is the most power-efficient constellation of this size for uncoded transmission, and subset-optimized PM-QPSK (SO-PM-QPSK) \cite{Sjodin2013_CL}. In terms of uncoded error probability, $\Cfs$ and SO-PM-QPSK offer asymptotic gains over PM-QPSK of $1.11$~dB and $0.44$~dB, resp. While $\Cfs$ is asymptotically the best constellation, PM-QPSK and SO-PM-QPSK have the advantage of a lower implementation complexity. The asymptotic gains offered by $\Cfs$ have been experimentally demonstrated in \cite{Karout2013_OFC, Bulow2013}, where it was also shown that $\Cfs$ gives higher MI than PM-QPSK at all SNRs. This indicates that $\Cfs$ is the best choice among these formats for capacity-approaching CM schemes with ML decoding. 

\begin{figure}
\newcommand{\scale}{0.8}
\newcommand{\scalesmall}{0.55}
\centering
\psfrag{xlabel}[cc][cB][\scale]{$\Eb/\No$~[dB]}%
\psfrag{ylabel}[cc][cb][\scale]{$\R$~[bit/symbol]}%
\psfrag{CAW}[cl][cl][\scalesmall]{Shannon cap.}%
\psfrag{PMQMI}[cl][cl][\scalesmall]{PM-QPSK}%
\psfrag{C416MI}[cl][cl][\scalesmall]{$\Cfs$}%
\psfrag{C416GMII}[cl][cl][\scalesmall]{$\Cfs$}%
\psfrag{SOPMQMI}[cl][cl][\scalesmall]{SO-PM-QPSK}%
\psfrag{SOPMQGMII}[cl][cl][\scalesmall]{SO-PM-QPSK}%
\includegraphics[width=0.48\textwidth]{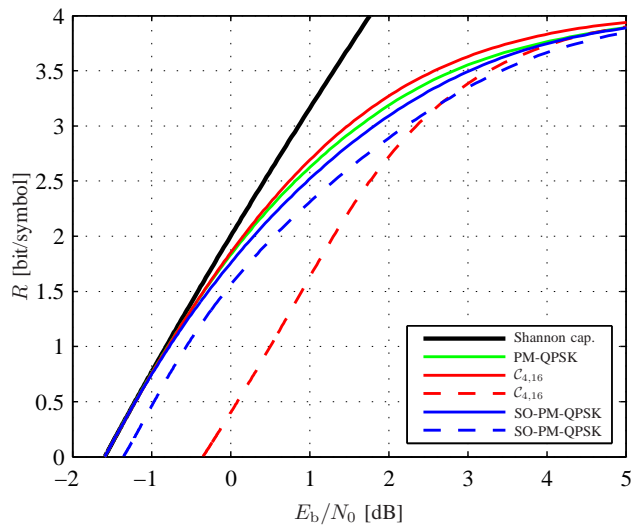}
\psfrag{xlabel}[cc][cB][\scale]{$\Es/\No$~[dB]}%
\psfrag{ylabel}[cc][cb][\scale]{BER}%
\psfrag{14}[cc][cc][\scale][-90]{{\fcolorbox{white}{white}{{$\Rc=1/4$}}}}%
\psfrag{12}[cc][cc][\scale][-90]{{\fcolorbox{white}{white}{{$\Rc=1/2$}}}}%
\psfrag{34}[cc][cc][\scale][-90]{{\fcolorbox{white}{white}{{$\Rc=3/4$}}}}%
\psfrag{91}[cc][cc][\scale][-90]{{\fcolorbox{white}{white}{{$\Rc=9/10$}}}}%
\includegraphics[width=0.48\textwidth]{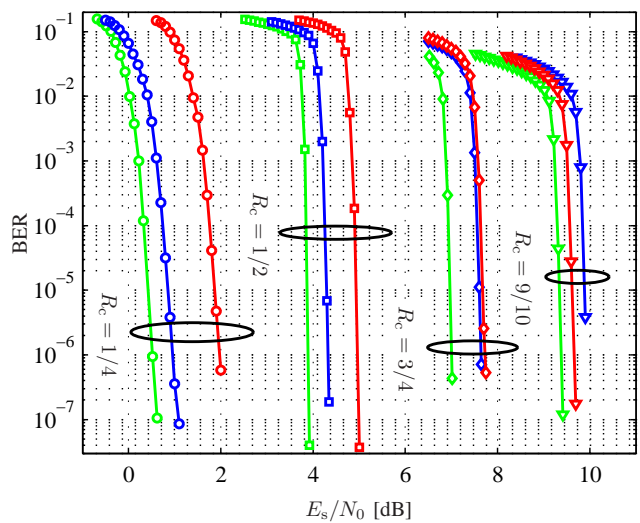}
\caption{Top: MI (solid lines) and GMI (dashed lines) for three constellations. Bottom: BER for the LDPC code with different rates and PM-QPSK (green), $\Cfs$ (red), and SO-PM-QPSK (blue).
}
\label{numerical_results}
\end{figure}

In \figref{numerical_results} (top), the MI and GMI for the three constellations under consideration are shown.\footnote{Calculated numerically using the ready-to-use Gauss--Hermite quadrature expressions in \cite[Sec.~III]{Alvarado11b}.}  For SO-PM-QPSK, we use the labeling proposed in \cite{Sjodin2013_CL}, while for $\Cfs$ we use a labeling (found numerically) that gives high GMI for a wide range of SNR. For PM-QPSK, we use the unique Gray code, which assigns a separate bit to each dimension. Thus, PM-QPSK becomes the Cartesian product of four 2-PAM constellations, and $\sum_{k=1}^{m}I(B_{k};\bY) = I(\bX;\bY)$. In other words, PM-QPSK causes no penalty if a bit-wise receiver is used. This is not the case for the two other constellations. The results in \figref{numerical_results} show that $\Cfs$ indeed gives a high MI at all SNRs; however, a large gap between the MI and GMI exists (more than $1$~dB for low rates). Therefore, $\Cfs$ will not work well with a bit-wise receiver. The situation is similar for SO-PM-QPSK, although in this case the losses are smaller. Interestingly, when comparing the GMIs for $\Cfs$ and SO-PM-QPSK, we observe that they cross at around $\R\approx 3.25$~bit/symbol. This indicates that a capacity-approaching scheme with a bit-wise receiver will perform better with $\Cfs$ than SO-PM-QPSK at high SNR. However, PM-QPSK is the best choice at any SNR.

To show that the conclusions above correspond to gains in terms of bit-error rate (BER), we simulated the three constellations with irregular repeat-accumulate LDPC codes. Each transmitted block consists of $64,800$ coded bits, four code rates $\Rc = 1/4, 1/2, 3/4, 9/10$ are considered, and the transmission rates are $\R=1.0,2.0,3.0,3.6$~bit/symbol, resp. The coded bits are assigned cyclically to the binary sequences $\un{B}_1, \un{B}_2, \un{B}_3, \un{B}_4, \un{B}_1, \ldots$, with no interleaver. At the receiver, LLRs are calculated and passed to the soft FEC decoder, which performs 50 iterations. The obtained BER results are shown in \figref{numerical_results} (bottom). Among the three constellations, PM-QPSK (green curves) always gives the lowest BER. The gains offered by PM-QPSK with respect to $\Cfs$ for low rates are about $1$~dB. More importantly, these gains are obtained by using a very simple demapper that computes four 2-PAM LLRs, one in each dimension. These results also show that the GMI curves in the top graph predict the coded performance of the system well. For example, the GMI curves indicate that at high coding rates, $\Cfs$ is better than SO-PM-QPSK, which is exactly what happens in terms of BER (\ie for $\Rc=9/10$, $\Cfs$ gives a lower BER than SO-PM-QPSK).

4D constellations with $M=256$ ($m=8$), \ie $2$~bit/dimension, were also investigated. In this case, we compared the GMI for two constellations: PM-16QAM, \ie a straightforward generalization of PM-QPSK formed as the Cartesian product of four 4-PAM constellations, and a numerically optimized constellation that gives low error probability at high SNR, which we denote by $\Cft$. The results obtained in this case are quite similar to the ones shown in \figref{numerical_results}, \ie the constellation $\Cft$ gives higher MI and lower GMI than PM-16QAM. Thus, $\Cft$ is unsuitable for a bit-wise receiver. A major advantage with PM-16QAM is the existence of Gray codes, which not only offer good performance but also let the LLRs be calculated in each dimension separately, thus reducing complexity.

\balance

\section{Conclusions}\label{Sec:Conclusions}

In this paper, we studied achievable rates for coherent optical coded modulation systems where the receiver is based on a bit-wise structure. Both analytical and numerical results show that simply transmitting and receiving independent data in each polarization is the best choice in this scenario. Multidimensional constellations, which are optimal with ML receivers and in uncoded systems, are not good for bit-wise receivers. On top of the weaker performance and higher decoder complexity, such constellation also carry the design challenge of selecting a good binary labeling.


\bibliographystyle{IEEEtran}
\bibliography{IEEEabrv,references_OFC_2013}

\end{document}